\documentclass[conference]{IEEEtran}
\IEEEoverridecommandlockouts

\usepackage{hyperref}
\usepackage{booktabs}
\usepackage[utf8x]{inputenc} 
\usepackage[]{caption}
\usepackage{listings}
\usepackage{multicol}
\usepackage[noabbrev,capitalize]{cleveref}
\crefname{section}{§}{§§}
\Crefname{section}{§}{§§}
\crefformat{section}{§#2#1#3}

\usepackage{tikz}
\usetikzlibrary{intersections}	
\usepackage{pgfplots}
\makeatletter
\@namedef{ver@transparent.sty}{}
\makeatother

\usepackage{multirow}
\usepackage{bold-extra}

\usepackage{makecell}
\usepackage{mathtools}
\usepackage{amsfonts}
\usepackage{booktabs}
\usepackage{amsthm}
\usepackage{tikz}
\usetikzlibrary{matrix}
\usepackage{multicol}
\usepackage{enumitem}
\usepackage{pifont}
\usepackage{amsmath}
\usepackage{graphics}
\usepackage{tikz}
\usepackage{colortbl}
\usepackage{pgfplots}
\pgfplotsset{compat=1.7}
\usepackage[ruled,vlined,linesnumbered,resetcount]{algorithm2e}
\SetAlFnt{\scriptsize\sffamily}
\SetKwComment{Comment}{$\triangleright$\ }{}
\SetEndCharOfAlgoLine{}

\SetCommentSty{mycommfont}
\definecolor{light-gray}{gray}{0.80}
\definecolor{darkgray}{rgb}{0.66, 0.66, 0.66}
\definecolor{gray}{rgb}{0.5, 0.5, 0.5}
\definecolor{alizarin}{rgb}{0.82, 0.1, 0.26}
\definecolor{amber}{rgb}{1.0, 0.75, 0.0}
\definecolor{english}{rgb}{0.0, 0.5, 0.0}
\definecolor{azure}{rgb}{0.0, 0.5, 1.0}
\definecolor{brass}{rgb}{0.71, 0.65, 0.26}
\definecolor{carnelian}{rgb}{0.7, 0.11, 0.11}
\pgfplotsset{width=7cm,compat=1.14}

\usepackage{numprint}
\usepackage{xspace}
\newcommand{\sysname}{{\small\textsc{Reckon}}\xspace}
\usepackage{paralist}

\renewcommand{\baselinestretch}{0.96}

\begin{document}
\pagestyle{plain}


\date{}

\title{Automated CFI Policy Assessment with Reckon}

\author{
{\rm Paul Muntean, Technical University of Munich, Germany}
} 

\maketitle

\begin{abstract}
Protecting programs against control-flow hijacking attacks recently has become an arms race between defenders and attackers. While certain defenses, e.g., \textit{Control Flow Integrity} (CFI), restrict the
targets of indirect control-flow transfers through static and dynamic analysis,
attackers could search the program for available gadgets that fall into the legitimate target sets
to bypass the defenses.
There are several tools helping both attackers in developing exploits and analysts in strengthening their defenses.
Yet, these tools fail to adequately
(1) model the deployed defenses, 
(2) compare them in a head-to-head way, and 
(3) use program semantic information to help craft the attack and the countermeasures.

Control Flow Integrity (CFI) has proved to be one of the promising defenses against control flow hijacks and tons of efforts have been made to improve CFI in various ways in the past decade. However, there is a lack of a systematic assessment of the existing CFI defenses. In this paper, we present \sysname, a static source code analysis tool for assessing state-of-the-art static CFI defenses, by first precisely modeling them and then evaluating them in a unified framework. \sysname helps determine the level of security offered by different CFI defenses, and find usable code gadgets even after the CFI defenses were applied, thus providing an important step towards successful exploits and stronger defenses. We have used \sysname to assess eight state-of-the-art static CFI defenses on real-world programs such as Google's Chrome and Apache Httpd.
\sysname provides precise measurements of the residual attack surfaces, and accordingly ranks CFI policies against each other. It also successfully paves the way to construct code reuse attacks and to eliminate the remaining attack surface, by disclosing calltargets under one of the most restrictive CFI defenses.

\end{abstract}


\section{Introduction}
\label{Introduction}  
Ever since the first Return Oriented Programing (ROP) attack \cite{rop:shacham}, the cat and mouse game between defenders and attackers has seen several peaks \cite{rop:carlini}.
As defenses improved over time, the attacks progressed with them \cite{rop:carlini}.
While defenders followed several lines of research when building defenses: 
control flow integrity \cite{zhang:vtrust, ropecker, perinput:niu, marx:tool, vci:asiaccs, vtint:zhang, vtv:tice, ivt, shrinkwrap, mcfi:niu, veen:typearmor}, 
binary re-randomization \cite{shuffler},
information hiding \cite{oxymoron}, and 
code pointer integrity \cite{cpi}, 
the attacks kept up the pace and got more and more sophisticated \cite{coop, coop:loop:oriented, subversive-c:lettner, blue:lotus, trap:crane}.

In principle, even with the myriad of currently available CFI defenses, performing exploits is still possible.
Attackers could search the program for gadgets that are allowed by CFI defenses 
to conduct \textit{Code Reuse Attacks} (CRAs) \cite{carlini:bending, coop}.
But the attacks become highly program-dependent,
and the applied CFI policies make reasoning about security harder. 
The attacker/analyst is thus confronted with the challenge of searching (manually or automatically) the protected program's binary or source code for gadgets which remain useful after CFI defenses have been deployed.
Thus, there is a growing demand for defense-aware assessing tools, that assist security analysts to assess CFI defenses.

%


Existing tools, including static pattern-based gadget searching tools \cite{ropdefender, multiarchitecture:wollgast} and dynamic attack construction tools \cite{newton, revery, jujutsu, losing:control:conti, bop}, all lack deeper knowledge of the protected program. 
As such, they can find CRA gadgets, but cannot determine if the gadgets are usable after a defense was deployed.
A recent work Newton \cite{newton} could assess and bypass deployed defenses,
but it relies on the real execution of the target program to gather sufficient information.

Consequently, with each applied defense, a more capable assessment tool needs ideally to: 
(1) model the defense as precisely as possible, 
(2) use program metadata in order not to solely rely on runtime memory constraints,
(3) use precise semantic knowledge about the protected program code, and
(4) provide absolute measurement numbers w.r.t. the remaining attack surface.
This allows to provide precise and reproducible measurements,
to decide which CFI defense is better suited for a given situation,
and to defend against or craft CRAs by searching available gadgets.

In this paper, we present \sysname, to our knowledge, the first static source code analysis framework for modeling and assessing static state-of-the-art CFI defenses w.r.t. the security level they offer and the remaining attack surface they have. It provides {\em a unified framework} to evaluate different CFI defenses, enabling a head-to-head comparison.
Note that, the use of different compilers or platforms would make the results of CFI evaluation incomparable. 
\sysname relies on the insight that, by carefully modeling a CFI defense into a comprehensive policy, the introduced constraints on call sites and call targets can be assessed during program compile time, by a unified compilation analysis component.

\sysname also provides a set of {\em expressive primitives},  which are able to characterize a wide range of static CFI policies. For example, \sysname provides static primitives related to types, class hierarchies and virtual table layouts.
These primitives could be used as building blocks to model many CFI policies. Further, \sysname provides the available/legitimate call targets under different CFI defenses. It can be reused by CRA attacks, \textit{e.g.,} the control flow bending attack as described by Carlini \textit{et al.} \cite{carlini:bending}, or be used to refine the analysis pipelines of existing attack construction or defense tools.

Note that, \sysname only focuses on assessing static CFI defenses, as these are more commonly deployed than dynamic defenses.
In addition, \sysname focuses on source code rather than on binary code,
as comparing various static CFI defenses against each other is more feasible in this way.
Moreover, the binary CFI policy implementations can be expressed precisely in source code.
Therefore, there is no need to look at the binary of the protected program since its source code provides
more semantic richness and precision w.r.t. the constraints imposed by each CFI defense.


We evaluate \sysname with common open source programs: NodeJS, Bind, Memcached, Httpd, Lighttpd, Nginx, Apache Traffic Server, Google's Chrome Web browser, and Redis. 
We show that \sysname can help the assessment of CFI defenses and is effective at finding gadgets, even with highly restrictive state-of-the-art CFI defenses deployed. 
In addition, we demonstrate how \sysname can be utilized to craft a code reuse attack. We also show how it can be effectively used to empirically measure the real attack surface reduction after a certain static CFI defense policy was used to harden a program's binary. Applications of \sysname go beyond CFI defense assessment framework, and we envision \sysname as a tool for defenders and software developers to highlight the residual attack surface of a program. 

In summary, we make the following contributions:
 
\begin{compactitem}

\item We present a static CFI attack model that is powerful and drastically lowers the bar for performing CRAs against state-of-the-art CFI defenses. With this model, we also introduce a new CFI defense metric ($CTR$) to more precisely characterize the existing CFI defenses. 

\item We implement \sysname, a novel framework usable for empirically analyzing and comparing CFI defenses against each other, as well as for generating low-effort CRAs by identifying the legitimate target sets under different CFI defenses and highlight how they can be used to craft attacks. \looseness=-1
 

\item We present general evaluation results based on several real-world programs by comparing existing static CFI defenses on multiple security relevant dimensions. Meanwhile, we also present a NodeJS-based case study with the goal of highlighting how \sysname can be used to craft CRAs against a state-of-the-art defense, \textit{e.g.,} the secure VTV/IFCC \cite{vtv:tice} implementation.


\end{compactitem}


\section{Background} 
\label{Background} 

\subsection{CRA Primitives}
There are mainly two types of CRA primitives: one exploits the forward-edge, and the other exploits the backward-edge, based on the program control flow graph (CFG).

\textbf{Forward-edge based CRAs} exploit the forward-edges of CFGs. 
First, at the source code level by performing calls through either function pointers
(\textit{e.g.,} single level of indirection) or virtual calls (\textit{e.g.,} double level of indirection). For example, these calls may use an array of function pointers that is accessed by a pointer to a virtual table (vtable) plus an index.
Second, at the binary level, \texttt{jump}, and \texttt{call} instructions are used to redirect the program control flow to a different address than the one intended in the original program CFG.
 
\textbf{Backward-edge based CRAs} violate CFG backward edges.
First, at the source code level, the function will not 
return to the next source code line from where the function was first called.
Second, in the program binary, the address located on the stack is modified such that the function's \texttt{ret} instruction will return to a different address than the one next to the instruction from where the function was initially called (mostly through a \texttt{call} instruction).

Finally, these two types of primitives (forward and backward edges) are used to link gadgets, in order to form a gadget chain with the goal of performing turing-complete malicious computations. Note that in this work we focus only on forward-edge based code reuse attacks and their assessment.

\subsection{Control-Flow Integrity}
Control-Flow Integrity (CFI) is a state-of-the-art technique used successfully along other techniques to protect forward and backward edges against program control-flow hijacking.
CFI is used to mitigate CRAs by, for example, pre-pending an indirect callsite with runtime checks that make sure only legal calltargets are allowed by an as precisely as possible  computed control flow graph (CFG) \cite{abadi:cfi2}. 

\smallskip\noindent
\textbf{Protection Schemes.}
Alias analysis in binary programs is undecidable \cite{alias:analysis}. For this reason, when protecting CFG forward-edges, defenders focus on using other program
primitives to enforce a precise CFG during runtime.
These primitives are most commonly represented by the program's: class hierarchy \cite{shrinkwrap}, virtual table layouts \cite{vfuard:aravind}, 
quasi-class hierarchies \cite{marx:tool}, binary function types \cite{veen:typearmor} (callsite/calltarget parameter count matching), etc.
They are used to enforce a CFG which is as close as possible to the original CFG being best described by the program control flow execution. 
Note that state-of-the-art CFI solutions use either static or dynamic information for determining legal calltargets.

\smallskip\noindent
\textbf{Static Information.}
CFI solutions that use static information allow callsites to target:
(1) all function entry points \textit{e.g.,} \cite{mingwei:sekar},
map callsite types to target function types by creating a mask which enforces that the number of provided parameters (up to six) has to be higher than the number of consumed parameters \textit{e.g.,} \cite{veen:typearmor},
(2) a quasi-class hierarchy (no root node(s) and the edges are not oriented) can be recuperated from the binary and enforced \textit{e.g.,} \cite{marx:tool},
(3) all virtual tables that can be recuperated and enforced \textit{e.g.,} \cite{vfuard:aravind},
only certain virtual table entries are allowed \textit{e.g.,} \cite{zhang:vtrust} based on a precise function type mapping, and
(4) sub-class hierarchies are enforced \textit{e.g.,} \cite{vtv:tice, shrinkwrap, ivt}.

\smallskip\noindent
\textbf{Dynamic Information.}
The goal of CFI solutions which use dynamic information is to refine their runtime analysis by leveraging program information which is only available during program execution.
In particular, PiCFI \cite{perinput:niu} restricts the set of calltargets to functions which have their address computed during runtime.
Context-sensitive solutions with different levels of context precision rely on hardware features such as the 
Last Branch Register (LBR) \cite{veen:cfi} to track a limited 
range (\textit{i.e.,} 16 up to 32 address pairs) of so called \textit{from} and \textit{to} addresses pairs during runtime. They then compare them against a precomputed program CFG. 
Finally, note that Intel Processor Trace (PT) \cite{griffin} can be used to build a longer history of address pairs compared to other approaches.


\section{Design of \sysname} 
\label{Overview} 
\subsection{Overview}

\sysname is designed as a static gadget detection framework which can be used to assist an analyst 
evaluating the attack surface after different types of static CFI defenses were applied.
To achieve this, \sysname applies a static black box strategy in order to statically retrieve code-reuse
gadgets through a set of attacker-controllable forward control flow graph (CFG) edges. The forward-vulnerable CFG edges are expressed as a callsite with a variable number 
of possible target functions. Further, these edges can be reused by an attacker to call arbitrary functions via arbitrary read or 
write primitives. To call such series of arbitrary functions, an attacker can chain a number of edges together by dispatching fake objects
contained in a vector. See, for example, the COOP \cite{coop} attack which is based on a dispatcher gadget used to call other gadgets through a single loop iteration. The COOP attack uses gadgets which are represented by whole virtual functions.

\sysname supports a wide range of code reuse defenses based on user-defined policies, which are composed of constraints about the set of possible calltargets allowed by a particular applied CFI defense.
The main idea behind \sysname is to compile the target program with different types of CFI policies 
and get the allowed target set per callsite for each constraint configuration. 
Note that we assume the program was compiled with the same compiler as the one on which \sysname is based.
Moreover, \sysname's policies are reusable and extensible; they model security invariants of important CRA defenses. Essentially, under these 
constraints, virtual pointers at callsites can be corrupted to call any function in the program. Thus, in this paper, we focus 
on the possibility to bend \cite{carlini:bending}
a pointer to the callsite's legitimate targets. Further, we assume that large programs contain enough gadgets for successfully performing CRAs. Bending assumes that it is possible for an attacker to reuse protected gadgets during an attack making the applied defense of questionable benefit.

\begin{figure}[ht]
\centering
   \includegraphics[width=.42\textwidth]{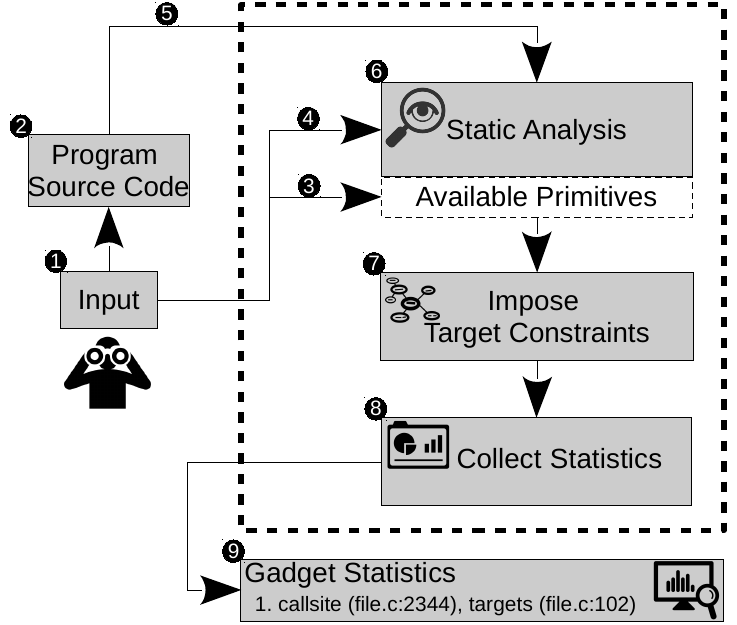}
    \caption{Design of \sysname.}
    \label{Design of CFI-Assessor.}
 \end{figure}
 
\autoref{Design of CFI-Assessor.} depicts the main components of \sysname and the workflow used for analyze the source code of a potentially vulnerable program in order to determine CRA statistics, as follows.
The analyst first provides as input \ding{182} to \sysname a program's source code \ding{183}.
Based on the previously selected defense, the desired analysis will be performed. The analysis previously mentioned is dependent on the selected defense model and on the available primitives. 
The analyst does this if he knows that the defense he wants to apply is currently available in the \sysname tool. The user selects the used defense model from the list of implemented 
defenses. This is done by switching on a flag inside the \sysname source code, which can also be implemented as a compiler flag, if desired.
In case the defense is not available, the analyst needs to extend the list of primitives \ding{184}, and model his defense as a policy (set of constraints) in the analysis
component of \sysname \ding{185}. 
In order to do this, he needs to know about the analysis internals of \sysname. 
After selecting/modeling a defense, the analyst forwards the application's source code \ding{186}  to \sysname which will analyze it with its static analysis component.
During static analysis \ding{187} the previously selected defense will be applied when compiling the program source code. As the analysis is performed, each callsite is constrained with only the legitimate calltargets. Note that the per callsite, a
legitimate calltarget is dependent on the currently selected defense model.
The result of the analysis contains information about the residual target set for each individual callsite after a CFI policy was assessed \ding{188}. This list contains 
a set of gadgets (callsites + calltargets) that can, given a certain defense model, be used to bend the control flow of the application. 
These target constraints are collected and clustered in the statistics collection component of \sysname \ding{189}.
Finally, at the end of the gadget collection phase, a list of calltargets containing potential usable gadgets statistics \ding{190} based on the currently applied defense(s) will be reported. 

\subsection{Reckon's Analysis Primitives}
\label{Available Primitives}
\sysname provides the following program primitives, which are either collected or constructed during program compile time. 
These primitives are used by \sysname to implement static CFI policies and to perform calltarget constraint analysis. 
Briefly, the currently available primitives are as follows:

\textbf{Virtual table hierarchy} (see \cite{shrinkwrap} for more detailed definition) allows performing virtual table inheritance analysis of only virtual classes as only 
 these have virtual tables. Finally, a class is virtual if it defines or inherits at least one virtual function.
 
\textbf{Vtable set} is a set of vtables corresponding to a single program class.
 This set is useful to derive the legitimate set of calltargets for a particular 
 callsite. The set of calltargets is determined by using the class inheritance relations contained inside a program.
 
\textbf{Class hierarchy} (see \cite{class:slicing, rossie:friedman} for a formal definition) can be represented as a class hierarchy graph with the goal to model inheritance relations between classes. Note that a real-world program can have multiple class hierarchies (\textit{e.g.,} Chrome, Google's Web browser). Note that the difference between virtual table hierarchy and class hierarchy is that the class hierarchy contains both virtual and non-virtual classes 
whereas the virtual table hierarchy can only be used to reason about the inheritance relations between virtual classes.
  
\textbf{Virtual table entries} allow \sysname to analyze the number of entries in each virtual table with the possibility to differentiate between
 virtual function entries, offsets in vtables, and thunks.

\textbf{Vtable type} is determined by the name of the vtable root for a given vtable. A vtable root is the last derived 
 vtable contained in the vtable hierarchy.
 
\textbf{Callsites} are used by \sysname to distinguish between direct and
 indirect (object-based dispatch and function-pointer based indirect transfers) callsites.

\textbf{Indirect callsites} are based on: (1) object dispatches or (2) function pointer based calls.
 Based on these primitives, \sysname can establish different types of relations between 
 callsites and calltargets (\textit{i.e.,} virtual functions). At the same time, we note that it is possible to derive 
 backwards relationships from calltargets to legitimate callsites based on this primitive.
 
\textbf{Callsite function types} allow to precisely determine the number and the type of the provided parameter by a callsite.
 As such, a precise mapping between callsites and calltargets is possible.

\textbf{Function types} allow to precisely determine the number 
 of parameters, their primitive types and return type value for a given function. 
This way, \sysname can generate a precise mapping between compatible
 calltargets and callsites.

These primitives can be used as building blocks during the various analyses that \sysname
can perform in order to derive precise measurements and a thorough assessment of a modeled static CFI policy.
We note that in order to model other CFI defenses, other (currently not available) simple or aggregated
analysis primitives may need to be added inside \sysname. \looseness=-1

\subsection{Constraints}
\label{Target Constraints Analysis}

The basic concept of any CRA is to divert the intended control flow of a program by using arbitrary memory write and read primitives.
As such, the result of such a corruption is to bend~\cite{carlini:bending} the control flow, such that it no longer points to  
the intended (legitimate) calltarget set. This means that 
the attacker can point to any memory address in the program. While this type of attack is still possible, we want to highlight another type of CRA
in which the attacker uses the intended/legitimate per callsite target set. That is, the attacker calls inside this set and performs his malicious behavior by reusing calltargets which are protected, yet usable during an attack. As previously observed by others \cite{newton}, CRA defenses try to mitigate this by mainly relying on one or two dimensions at a time, as follows:

 \textbf{\textit{Write constraints}} limit the attacker's capabilities to corrupt writable memory. If there is no defense in place, 
 the attacker can essentially corrupt: pointers to data, non-pointer values such as strings, and pointers to code (\textit{i.e.,} 
 function pointers). In this paper, we do not investigate these types of defenses as these were already addressed in detail by Veen \textit{et al.} \cite{newton}. 
 Instead, we focus on target constraints as these represent a big class of defenses which in our assessment needs separate and detailed 
 treatment. This obviously does not mean that our analysis results cannot be used in conjunction with dynamic write constraint 
 assessing tools. Rather, our results represent a common ground truth on which runtime assessing tools can build their gadget detection analysis.
 
 \textit{\textbf{Target constraints}} restrict the legitimate calltarget set for a callsite which can be controlled by an attacker.
 With no target constraints in place, the target set for each callsite is represented by all functions located in the program and any linked shared library.
 The key idea is to reduce the wiggle room for the attacker such that he cannot target random callsites. 
 As most of these defenses impose a one-to-$N$ mapping, an attacker being aware of said mapping could corrupt the pointer at the callsite to bend \cite{carlini:bending}
 the control flow to legitimate targets
 in an illegitate
 order to achieve his malicious goals. This essentially means that all static defenses impose target constraints.

\textbf{Static Analysis.} 
\sysname is based on the static analysis of the program which is represented in LLVM's intermediate representation (IR). The analysis is performed 
during link time optimization (LTO) inside the LLVM \cite{llvm:framework} compiler framework to detect callsites and legitimate callees under the currently analyzed CFI defense. 
\sysname uses the currently available primitives and the implemented defenses to impose target constraints for each callsite individually. 
Currently, 8 target defenses are supported, see \cref{Implementation}, but this list can easily be extended since 
all defenses are based on the similar mechanisms which assessable during a whole program analysis. 

\textbf{Generic Target Constraints.} 
As mentioned above, \sysname can be used to impose existing generic calltarget constraints (defenses) based on class hierarchy relations and 
callsites and calltarget type matching with different levels of precision depending on the currently modeled CFI defense. 
Further, \sysname allows 
extending and combining existing policies or applying them concurrently.


\section{Implementation} 
\label{Design} 

\subsection{Data Collection and Aggregation}
\textbf{Collection.} \sysname collects the virtual tables of a program in the Clang front-end and pushes them through the compilation pipeline in 
order to make them available during link-time optimization (LTO). For each virtual table, \sysname collects the number of entries.
The virtual tables are analyzed and aggregated to virtual table hierarchies in a later step.
Other data such as direct/indirect callsites and function signatures
are collected during LTO. 

\textbf{Aggregation.} Next we present the program primitives which are constructed by \sysname:
(1) virtual table hierarchies based on the previously collected virtual 
tables inside the Clang front-end. The virtual table hierarchies are used to derive relationships between the classes 
inside a program (class hierarchies), determine sub-hierarchy relationships and count, for example, how many virtual table entries (virtual functions)
a certain virtual table sub-hierarchy has. 
(2) virtual table sets which are used for mapping callsites to legitimate class hierarchy-based virtual calltargets. 
(3) callsite function types which are composed of the number of parameters provided by a callsite, their types, and if the callsite is a void or non-void callsite. 
(4) function types which are composed of the function name, the expected number of parameters and their types and if the function is a void or non-void function.

\vspace{-0.1in}
\subsection{CFI Defense Modeling}
\sysname implements a set of constraints for each modeled CFI-defense, which are defined as analysis conditions that model the behavior of each analyzed CFI-defense.
These constraints are particular for each CFI-defense and operate on different primitives. More specifically, different constraints of a CFI-defense
are implemented inside \sysname. The steps for modeling a CFI defense are as follows:
 (1) Which \sysname's primitives are used by the policy?
 (2) Is there a nesting or subset relation between primitives?
 (3) Does the policy rely on hierarchical meta-data primitives?
 (4) What is the callsite/calltarget matching criteria?
 (5) How to count a callsite/calltarget match?

Note that there is no effort needed to port \sysname from one policy to another as all policies can 
operate in parallel during compile time. As such, the measurement results obtained for each policy are written 
in one pass in an external file for later analysis.

Next, we give a concrete example of how a CFI defense, TypeArmor's \textit{Bin types} policy \cite{veen:typearmor}, was modeled inside \sysname by following the steps mentioned above. For more details, see Section \ref{typearmor:policy} for a description on how this policy works and the constraint sub-classes depicted in Section \ref{Constraint Subclasses}. More specifically, for TypeArmor,
 (1) The policy uses the: callsite primitive, indirect callsite, callsite function type, and function type primitives provided by \sysname.
 (2) From all functions contained in the program, we analyze only the virtual functions which expect up to six parameters to be passed by the callsite. 
 Next from all callsites, we filter out the ones which are not calling virtual functions and which provide more than six parameters to the calltarget. 
 Check if the callsite is a void or non-void callsite. Check if each
 analyzed calltarget is a void or non-void target.
 (3) The policy does not rely on hierarchical meta-data.
 (4) A callsite matches a calltarget if it provides less or the same number of parameters as the calltarget expects.
 (5) In case the matching criteria holds, we count the total count one up for each found match. \looseness=-1

Finally, these constraints are implemented as an LLVM compiler module pass performed during LTO. Note that even an analyst with restricted previous knowledge can model the constraints
of a CFI policy by observing how other existing policies were implemented inside \sysname.

\subsection{CFI Defense Analysis}
\sysname performs for each implemented CFI defense a different analysis. Each defense analysis consists of one or more iterations through the program primitives which are relevant for 
the CFI defense currently being analyzed. 
Depending on the particularities of a defense, \sysname uses different previously collected program
primitives. More specifically, class hierarchies, class sub-hierarchies, or function signatures located in the whole program or in certain class sub-hierarchy are individually analyzed.
During a CFI-defense analysis, statistics are collected w.r.t. the number of allowed calltargets per callsite taking into account the previously modeled CFI-defense.

For example, for a certain CFI defense (\textit{e.g.,} TypeArmor's CFI policy \textbf{Bin types}) it is required to determine a match between the number of 
provided parameters (up to six parameters) of each indirect callsite 
and all virtual functions present in the program (object inheritance is not taken into account) which could be the target (may consume up to six parameters) of such a callsite. 
In order to analyze this CFI defense and collect the statistics, \sysname visits all
indirect callsites 
it previously detected in the program and all virtual functions located in all previously recuperated class hierarchies.
Afterwards, each callsite is matched with potential calltargets 
(virtual functions).  
Finally, after all virtual callsites and virtual functions were visited, 
the generated information is presented to the analyst.

\subsection{Implementation Details}
We implemented \sysname as one link time optimization (LTO) pass
by extending the Clang/LLVM (v.3.7.0) compiler~\cite{clang:llvm} framework infrastructure. 
The implementation of \sysname is split between the Clang compiler front-end (part of the metadata is  collected here), and one 
link-time pass, totaling around 4.2 KLOC.
\sysname supports separate compilation by relying on the LTO mechanism built in LLVM~\cite{clang:llvm}. 
By using Clang, \sysname collects  front-end virtual tables and makes them available during LTO. 
Next, virtual table hierarchies are built which are used to model different CFI defenses.
Other \sysname primitives such as function types are constructed during LTO. 
Finally, each of the analyzed CFI defense is separately modeled inside \sysname by using 
the previously collected primitives and aggregated data to impose the required defense constraints.

\section{Analyzing CFI Defenses with \sysname} 
\label{Mapping Defenses Into CFI-Assessor} 
In this section, we show how to map real-world CFI defenses into \sysname based on available primitives.
Similarly to Newton
\cite{newton}, \sysname models the security provided by CRA defenses along two axes: (1) write constraints imposed by the defense, and (2) imposed target constraints.
The main motivation for the usage of this modeling technique is that: (1) the majority of CFI techniques do not impose write constraints, with the exception of CFI runtime tools which adjust their analysis by using hardware features, and
(2) it serves as a natural extension (\textit{i.e.}, it eases understanding) of existing work.
Next, we will map the defenses presented in \cref{Background} according to these constraints. 
In our work, we focus only on one class of defense at a time and 
we add more constraint types to the target constraint axis. As opposed to Newton \cite{newton}, \sysname helps to derive static constraints imposed on the target program.  
This mapping allows to define textual descriptions of each CFI policy and reduces the task to a compiler-based counting problem allowing an analyst to determine which callsites and 
calltargets are protected and which are not.

\subsection{Deriving Constraints}
\begin{table}[ht]
\centering
\resizebox{0.99\columnwidth}{!}{%
 \begin{tabular}{ l | l | l | c | c | l }
                            \multicolumn{3}{c}{\hspace{1.7cm}\textbf{Defense}}                                                            & \multicolumn{1}{c}{\textbf{Write Constraints}}  & \multicolumn{2}{c}{\textbf{Target Constraints}}\\
          \textit{Class}    & \textit{Subclass}     & \textit{Solutions}                                                                  & \textit{Details}  & \textit{Dynamic}  &\textit{Details}\\ \hline
                            &TypeArmor              & \cite{veen:typearmor}                                                               & None              & \ding{56}         & Bin types             \\
                            &IFCC/MCFI              & \cite{mcfi:niu, vtv:tice}                                                           & None              & \ding{56}         & Src types             \\
                            &Safe IFCC/MCFI         & \cite{mcfi:niu, vtv:tice}                                                           & None              & \ding{56}         & Safe src Types        \\
                            &ShrinkWrap/IVT         & \cite{shrinkwrap}                                                                   & None              & \ding{56}         & Strict sub-hier.   \\
              CFI           &VTV                    & \cite{vtv:tice, ivt}                                                                & None              & \ding{56}         & Sub-hierarchy         \\
                            &vTint                  & \cite{vtint:zhang}                                                                  & None              & \ding{56}         & All vTables           \\
                            &Marx/VCI               & \cite{marx:tool, vci:asiaccs}                                                       & None              & \ding{56}         & vTable hierarchy      \\
                            &HCFI                   & \cite{perinput:niu}                                                                 & None              & \checkmark        & Computed              \\
                            &CsCFI                  & \cite{ropecker}                                                                     & Segr              & \checkmark        & None                  \\
                            &vTrust                 & \cite{zhang:vtrust}                                                                 & None              & \ding{56}         & Strict src types     \\
\end{tabular}
}

\caption{Mapping of CFI code-reuse defenses into \sysname constraints. Note that all these CFI defenses were published in top tier security conferences.}
\label{Mapping of code-reuse defenses to constraints.}
\end{table}
\autoref{Mapping of code-reuse defenses to constraints.} highlights the constraints imposed by each defense subclass. In our work, we do not consider runtime-based write constraining defenses (HCFI and CsCFI, see Newton \cite{newton} for more details), since these runtime constraints are hard to be assessed statically. Instead, \sysname models a comprehensive set of eight defense classes with no write constraints, which we discuss in detail in the next subsection.  \looseness=-1

\subsubsection{Constraint Subclasses}
\label{Constraint Subclasses}
\textsc{}
\sysname provides the following constraint subclasses.
Note that the next three constraint classes (highlighted in bold italic font) were first presented by Newton \cite{newton}, while the last five classes are presented for the first time in this work.

\textbf{\textit{TypeArmor}} imposes callsite target constraints based on a function type policy. In particular,
for each callsite only targets which fulfill the parameter count based policy are allowed during 
runtime. 

\textbf{\textit{IFCC/MCFI}} enforces similar constraints as TypeArmor, with the exception that the function type is computed at the source rather than at the binary level.

\textbf{\textit{Safe IFCC/MCFI}} contains the same defenses as the IFCC/MCFI defense subclass, except that in this case, we distinguish a \textit{safe} mode, 
where type information is less strict for compatibility reasons with real-world programs, this is discussed in the original IFCC paper \cite{vtv:tice}.



\textbf{\textit{VTV}} subclass enforces for each indirect callsite the whole subclass hierarchy \textit{Src sub-hierarchy} which 
is precise but leaves wiggle room for the attacker and 
it enforces too many calltargets as noted in \cite{shrinkwrap}. 

\textbf{\textit{ShrinkWrap/IVT}} subclass can enforce a more precise class sub-hierarchy (\textit{Strict src types}) than IFCC/MCFI. 
For each indirect callsite (object dispatch) a precise class hierarchy is computed.

\textbf{\textit{vTint}} subclass operates at the binary level in order to find indirect callsites and virtual tables. 
This subclass is enforcing for all detected virtual tables (\textit{all vtables}) all its entries for each callsite.

\textbf{\textit{Marx/VCI}} subclass operates on binary to recuperate a quasi class hierarchy: 
(1) the class has no root node, and
(2) the edges in the class hierarchy are not oriented. 

\textbf{\textit{vTrust}} subclass enforces matching function types for each indirect callsite. It follows a precise source code callsite-to-targets mapping (\textit{Strict src types}) based on computing at all calltargets a hash composed of the number, types, and return type of each virtual function. 
Next, we use an example to show how the calltarget sets differ depending
on the used CFI defense.


\begin{center} \centering
 \begin{lstlisting}[caption={C++ class hierarchy with four classes.}, 
captionpos=b, 
label={vtrust_vs_cha_policy}, 
basicstyle=\footnotesize]
1.class Foo { virtual void get(); };
2.class Bar:public Foo { virtual void get(); }
3.class Baz:public Bar { virtual void get(); }
4.class Bac:public Bar { virtual void get(); 
5.                       virtual void set(); }
6.Bar *b = new Bar();
7.b->get();
\end{lstlisting}
\end{center}

\autoref{vtrust_vs_cha_policy} shows a simple \verb!C++! class hierarchy which will be used to show that
based on the used CFI policy different
functions are accessible after applying a certain policy.
For example, under the vTrust CFI policy the indirect call, line seven in \autoref{vtrust_vs_cha_policy} can target any of the four \texttt{get()} functions located in the \texttt{Foo}, \texttt{Bar}, \texttt{Baz}, and \texttt{Bac} since the function signatures match since its policy includes function names as well. For example, vTint allows all four \texttt{get()} functions and the \texttt{set()} function as targets, as it allows all entries located in all virtual tables detected in the program. Note that at the binary level vTint cannot distinguish between function name as the information vanishes through compilation.
Further, under the class hierarchy based analysis (CHA) of IVT, \texttt{Bar::get()}, \texttt{Bac::get()}, \texttt{Baz::get()} and \texttt{Bac::set()} can be called,
since this policy enforces all class sub-hierarchies in the protected program.

Shrinkwrap's CHA policy allows only \texttt{Bar::get()} and \texttt{Baz::get()} to be called as this policy is aware of primary and secondary inheritance paths inside the program virtual table hierarchy. Note that this holds since \texttt{Baz} class is considered a primary type while the \texttt{Bac} class is a secondary type in this virtual table hierarchy. Further, both ShrinkWrap and IVT policies are based on the type of the object. As such, only some subtypes are allowed. vTrust, for example, relies on the base objects and is therefore less precise than CHA-based policies. IVT, for example, allows all subtypes. Note that with a larger class hierarchy vTrust and IVT policies would allow the same number of targets are allowed. In contrast, ShrinkWrap allows just a sub-part of the class sub-hierarchy (some subtypes).

Finally, these CFI policy based target set constrining examples highlight the fact that the CFI policies have different granularities w.r.t. constraining the calltarget set per callsite.





\subsection{Describing and Analyzing CFI Defenses}
\label{Implementation}

\begin{figure}[ht]
\centering
\captionsetup{singlelinecheck = false, justification = justified}
   \includegraphics[width=.79\columnwidth]{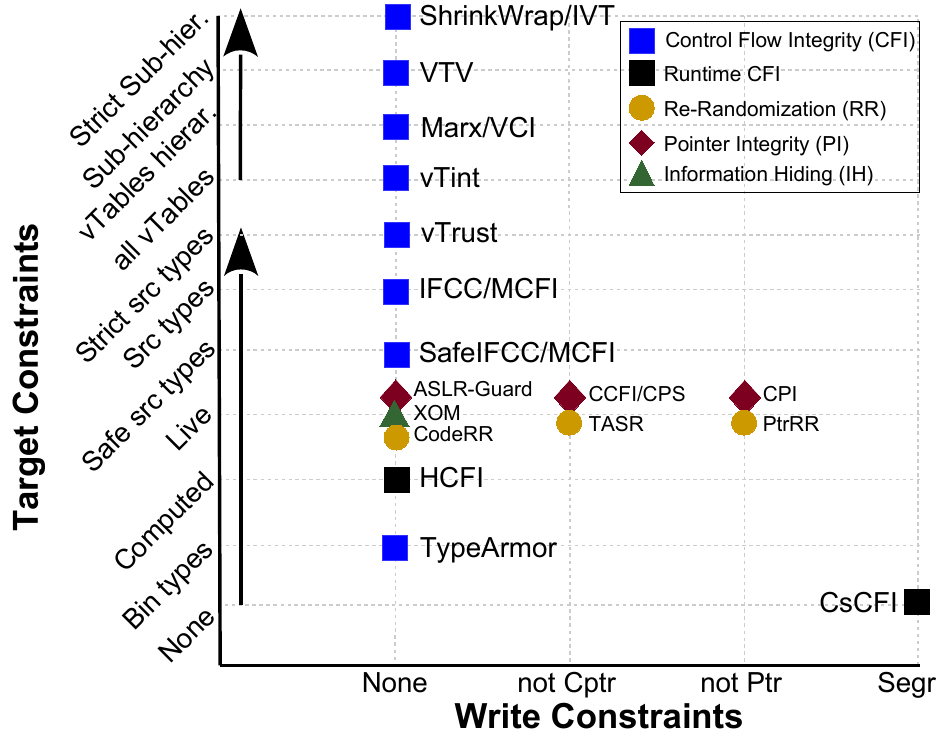}
    \caption{Mapping of CFI defense classes to write ($X$ axis) and target ($Y$ axis) constraints in \sysname.
    }
    \label{mapping}
\end{figure}
\autoref{mapping} depicts the constraints of the defenses, presented in 
\autoref{Mapping of code-reuse defenses to constraints.}. 
The constraints are grouped in two categories in ascending order indicated with
two arrows on the $X$ axis according to their precision
of determining precise target sets per callsite.
The $X$ axis shows the 
write constraints imposed by each defense subclass, and the $Y$ axis 
shows the target constraints. Defenses that share both the same write and 
target constraints impose equivalent security restrictions, thus each ($X$ axis, $Y$ axis)
pair depicted in \autoref{mapping} represents an equivalence class.

Note that in this paper, we are interested in only static CFI defenses which do not impose
any write constraints. We are aware that these defenses are thought to be accompanied
by some type of write constraint but for the sake of simplicity, we abstract 
these away for now. 
As such, runtime CFI (black rectangle), re-randomization (yellow circle), 
pointer integrity (brown diamond), and information hiding (green triangle) depicted 
in \autoref{mapping} are not taken into consideration as these were thoroughly analyzed
by Newton \cite{newton}. 
Our objective is to focus on the details of the static CFI class in order to 
reveal novel insights of this widely used class of defense.
marx:tool
\sysname can compile a given program with DWARF \cite{dwarf} information. In this situation, 
the function name and location inside the program is back-traceable to the exact
location in the source file. The same level of detail is possible for the associated callsite. 

Important to observe is that all static CFI defenses impose no write constraints
and as such it is recommendable to use these defenses with some type of write constraints which for
example would impose that the virtual pointer used dur\cite{vtv:tice, ivt}ing object dispatches cannot be overwritten
to point to an illegitimate address inside the program. 
Further, the value of this pointer is determined during runtime.
We further assume that there are no implementation specific vulnerabilities of these defenses
and as such the equivalence classes hold.
Further, our constraint-based classification abstracts away specific implementation details
and ignores implementation specific differences across these defenses.
For example, we consider the used primitives (\textit{e.g.,} class hierarchy) for enforcing the 
CFI policy to be ideal with no precision differences between binary and source code based tools.
As such, we are able to generate lower bounds w.r.t. the legitimate calltarget set
of these defenses and further, our approach is applicable to general gadget generation 
constraints across many different defenses.
 
Next, we show how to implement in \sysname the required constraints for the eight assessed defenses, by 
using the primitives specified in Section \ref{Available Primitives}. 
Specifically, we present in detail the constraints
imposed at any indirect callsite depending on the target constraint.

\textbf{Corrupting code pointers.} The CFI defenses which do not impose any write constraints are allowing any memory to be corrupted,
such as code pointers. This is presented in \autoref{mapping} by abstracting the CFI constraints to a counting problem of legitimate calltargets per callsite. 
These defenses are located on the left hand side of \autoref{mapping} depicted on the $X$ axis with (\textit{None}).
Note that in general, these type of constraints assume that the callsite is not corruptible since it is located in read-only memory, but since these constraints do not impose any specific code pointer corrupting defense these are regarded in this paper as not defending 
code pointer at all.  \looseness=-1

The particular constraint-based counts are obtained by computing for each constraint the target set counts with our \sysname tool. 
Next, \sysname compiles any given program in order to obtain the target set and performs counts according to the currently used constraint.
Note that while in \cref{Constraint Subclasses}, we provide the general description of how the constraint classes work, in this section, we
describe how the target constraints work in detail by associating them to the \sysname primitives.

In the following, we will present eight static CFI defense classes modeled inside \sysname. 
These defenses are stemming from published research papers and are used to constrain forward edge program control flow transfers to point to only legitimate calltargets.
Note that each CFI defense description is an idealized and very close to how the original CFI defense 
policy which was implemented in each tool. Finally, note that each modeled defense was previously thoroughly discussed with the original authors and only after the authors agreed with these descriptions we modeled them into \sysname. Next we give the formal definitions of 
each of the CFI defenses as these were modeled inside \sysname and the description of the performed analysis. 

\textbf{Notation.} The used notation is as follows: 
$P$ represents the analyzed program,
$Cs$ is the set of program indirect callsites (virtual and non-virtual),
$Cs_{virt}$ is the set of program virtual callsites,
$V$ is a virtual table hierarchy, 
$V_{sub}$ is a virtual table sub-hierarchy, 
$v_t$ is a virtual table,
$v_e$ is a virtual table entry (virtual function), 
$vc_s$ is a virtual callsite,
$nv_f$ is a non-virtual function, 
$v_f$ is a virtual function (virtual table entry), 
$C$ is a program class hierarchy,
$C_{sub}$ is a program class sub-hierarchy,
$c_s$ is a indirect callsite (object dispatch or indirect pointer based function call), 
$nt_{pcs}$ represents the number and type of parameters provided by a callsite, 
$nt_{pct}$ represents the number and type of parameters provided by a calltarget, 
$F$ set of all virtual and non-virtual functions contained in the program,
$S$ is the set of function signatures, 
$M$ is the policy matching set of rules. Note that $M$ is determined by all rules defined by a CFI defense and represents, at the same time, the matching criteria for each policy. This means that \sysname increments the count of his analysis by one when such a match is found.

  \textbf{Bin Types. (TypeArmor) \cite{veen:typearmor}} 
  \label{typearmor:policy} 
  We formalize this policy $\psi$ as the tuple   
  $\big \langle Cs, F, V, M \big \rangle$ where the relations hold:
  (1) $V \subseteq F$,
  (2) $v_e \in V$,
  (3) $nv_f \in F$,
  (4) $c_s \in C$, and
  (5) $M \subseteq Cs \times V \times F$.
  
  
  
   
  \textit{\sysname's Analysis.} For each indirect callsite $c_s$ (1) count the total number of virtual table entries $v_e$ which reside in each virtual table hierarchy $V$ contained in program $P$, and also, (2) count the number of non-virtual functions $nv_f$ residing in $F$, which need at most as many function parameters as provided by the callsite and up to six parameters. 
  Further, if $F$ contains multiple distinct virtual table hierarchies (islands) then continue to count them too and take them also into consideration for a particular callsite. 
  An island is a virtual table hierarchy which has no father child relation to another virtual table hierarchy contained in the program $P$.
 
  \textbf{Safe src types. (Safe IFCC) \cite{vtv:tice}} 
    We formalize this policy $\psi$ as the tuple   
  $\big \langle Cs, F, F_{virt} S, M \big \rangle$ where the relations hold:
  (1) $V \subseteq F$,
  (2) $v_f \in F_{virt}$,
  (3) $nv_f \in F$,
  (4) $nt_{pcs} \in S$,
  (5) $nt_{pct} \in S$,
  (6) $f_{rt} \in S$,
  (7) $c_s \in Cs$, and
  (8) $M \subseteq Cs \times F \times S$.
  
    \textit{\sysname's Analysis.} {For each indirect callsite $c_s$ count the number of virtual functions $v_f$ and non-virtual functions $nv_f$ located in the program $P$ for which the number and type of parameters required by the calltarget $nt_pct$ matches with the number and type of parameters provided at the callsite $nt_pcs$. 
 The function return type $f_{rt}$ of the matching function is not taken into consideration. 
 All parameter pointer types are considered 
 interchangeable, \textit{e.g.,} \textbf{int*} and \textbf{void*} pointers are considered interchangeable.}

  \textbf{Src types. (IFCC/MCFI) \cite{mcfi:niu}} 
   We formalize this policy $\psi$ as the tuple   
  $\big \langle Cs, F, F_{virt} S, M \big \rangle$ where the relations hold:
  (1) $V \subseteq F$,
  (2) $v_f \in F_{virt}$,
  (3) $nv_f \in F$,
  (4) $nt_{pcs} \in S$,
  (5) $nt_{pct} \in S$,
  (6) $f_{rt} \in S$, 
  (7) $c_s \in Cs$, and 
  (8) $M \subseteq Cs \times F \times S$.
  
    \textit{\sysname's Analysis.} {For each indirect callsite $c_s$ count the number of virtual functions and non-virtual functions located in the program $F$
 for which the number and type of parameters required at the calltarget $nt_{pct}$ matches the number and type of arguments provided by the callsite $nt_{pcs}$. 
 The return type of the matching function is ignored. Compared to \textit{Safe src types} this policy distinguishes between different pointer types, this means that these are not interchangeable and that the function signatures are more strict.
 Neither the return value of the matching function nor the name of the function are taken into consideration.}
 
 \textbf{Strict src types. (vTrust) \cite{zhang:vtrust}} 
  We formalize this policy $\psi$ as the tuple   
  $\big \langle Cs, V, F, F_{virt}, S, M \big \rangle$ where the relations hold:
  (1) $V \subseteq F$,
  (2) $v_f \in F_{virt}$,
  (3) $ntf_{pcs} \in S$,
  (4) $f_{rt} \in S$, 
  (5) $c_s \in Cs$, and 
  (6) $M \subseteq Cs \times S \times F_{virt} \times V$.
  
  \textit{Performed Analysis.} {For each indirect callsite $c_s$ compute the function signature of the function called at this particular callsite using the number of parameters, their types, and the name of the function $ntf_{pcs}$ (the literal name used in \verb!C/C++! without any class information attached). Match this function type identifier with each virtual function $v_f$
contained in each virtual table hierarchy $V$ of $P$. The name of the function is taken into consideration when building the hash
but not the function return type $f_{rt}$ as this can be polymorphic. Note that we have a match when the signature of a function called by a callsite matches with the signature of a virtual function $v_f$.}
 
 \textbf{All vtables. (vTint) \cite{vtint:zhang}} 
  We formalize this policy $\psi$ as the tuple   
  $\big \langle P, Cs, F_{virt}, V, M \big \rangle$ where the relations hold:
  (1) $V \subseteq F$,
  (2) $v_e \in V$,
  (3) $v_f \in F_{virt}$,
  (4) $c_s \in Cs$, and 
  (5) $M \subseteq Cs \times V$.
  
   \textit{\sysname's Analysis.} {For each indirect callsite $cs$ count each virtual function $v_f$ corresponding to a virtual table entry $v_e$ contained in each virtual table present in the program $P$.}
 
 \textbf{vTable hierarchy/island. (Marx) \cite{marx:tool}} 
  We formalize this policy $\psi$ as the tuple   
  $\big \langle P, F_{virt}, C, Cs, V, M \big \rangle$ where the relations hold:
  (1) $V \subseteq F$,
  (2) $v_e \in V$,
  (3) $v_f \in F_{virt}$,
  (4) $v_t \in V$,
  (5) $V \in C$,
  (6) $C \in P$,
  (7) $c_s \in Cs$, and
  (8) $M \subseteq Cs \times V \times C$.
  
  \textit{\sysname's Analysis.} {For each indirect callsite $c_s$ count each virtual function $v_f$ corresponding to each virtual table $v_t$ entry $v_e$ having the same index in 
 the virtual table as the index determined at the callsite $c_s$ by Marx.
 Perform this matching for each virtual table $v_t$ where the index matches with the index determined at the callsite $c_s$ and which is located in the class hierarchy $C$ which contains the class type of the dispatched object.
 Note that abstract classes are not taken in consideration within this policy, this can be recognized though by virtual tables having pure virtual function entries.}  \looseness=-1

  \textbf{Sub-hierarchy. (VTV) \cite{vtv:tice}} 
   We formalize this policy $\psi$ as the tuple   
  $\big \langle P, F_{virt}, C, C_{sub}, V, M \big \rangle$ where the relations hold:
  (1) $v_t \in V$,
  (2) $V \subseteq C$,
  (3) $C \subseteq P$,
  (4) $C_{sub} \in C$,
  (5) $v_f \in F_{virt}$,
  (6) $vc_s \in P$, and 
  (7) $M \subseteq Cs_{virt} \times C_{sub} \times V \times F_{virt}$.
  
    \textit{\sysname's Analysis.} {For each virtual callsite $vc_s$ build the class sub-hierarchy $C_{sub}$ having as root node the base class 
  (least derived class that the dispatched object can be of) of the dispatched object. From the classes located in the sub-hierarchy consider, for the currently analyzed callsite,
each virtual table $v_t$. Further within
this virtual tables $v_t$'s consider only the virtual function $v_f$ entries located at the offset used by the virtual object dispatch mechanism. Next count the number of virtual functions to which these entries point to.}

 \textbf{Strict sub-hierarchy. (ShrinkWrap) \cite{shrinkwrap}} 
  We formalize this policy $\psi$ as the tuple   
  $\big \langle P, F_{virt},  C, V, V_{sub}, M \big \rangle$ where the relations hold:
  (1) $V \subseteq C$,
  (2) $v_e \in V$,
  (3) $v_f \in F_{virt}$,
  (4) $v_t \in V$,
  (5) $V \subseteq C$,
  (6) $V_{sub} \subseteq V$,
  (7) $C \subseteq P$,
  (8) $c_s \in P$, and
  (9) $M \subseteq Cs_{virt} V \times V_{sub} \times F_{virt}$.
  
   \textit{\sysname's Analysis.} {For each virtual callsite $vc_s$ identify the virtual table $v_t$ type used.
 Take this virtual table $v_t$ from the base class $C$ of the dispatched object and build the virtual 
 table $v_t$ sub-hierarchy $V_{sub}$ having this virtual table $v_t$ as root node.
 From the virtual tables in this $v_t$ sub-hierarchy find the virtual function $v_f$ entries located at the offset used by the virtual object dispatch mechanism for this particular callsite $c_s$. Next count each virtual function $v_f$, to which these virtual table entries $v_e$ point to.} Finally, after \sysname computes for each callsite the total calltarget set count, as above described for each policy, it sums up all results for each callsite to generate several statistics.  \looseness=-1

\section{Evaluation}
\label{Evaluation}
\begin{table*}[ht!]
\centering
\resizebox{2.05\columnwidth}{!}{%
 \begin{tabular}{ l | r | r | r | r | r | r | r | r | r | r | r} \hline
         \multicolumn{1}{c|}{}                                          & \multicolumn{5}{c|}{\hspace{.3cm}\textbf{Targets Median}}                            & \multicolumn{6}{c}{\hspace{.2cm}\textbf{Targets Distribution}}\\
         \multicolumn{1}{l|}{\textbf{\textit{P}}}           & \multicolumn{5}{c|}{\hspace{.3cm}\textbf{}}                                          & \multicolumn{3}{c|}{\hspace{.2cm}\textbf{NodeJs}}  & \multicolumn{3}{c}{\hspace{.2cm}\textbf{MKSnapshot}}\\
         {\textbf{\textit{}}}                                 & \textit{NodeJS} & \textit{MKSnaphot} & \textit{Total} & \textit{Min} & \textit{Max}  & \textit{Min}  &\textit{90p}   &\textit{Max}   & \textit{Min}  &\textit{90p}   &\textit{Max}       \\ \hline
         {(1)}                                & 21,950 (21,950) & 15,817 (15,817)    & 15,817 (20,253)  & 15,817 (15,817)& 21,950 (21,950) & 12,545 (885)   & 30,179 (30,179) & 32,478 (32,478) & 8,714 (244)    & 21,785 (21,785) & 23,376 (23,376)  \\
         {(2)}                                 & 2,885 (88)      & 2,273 (495)        & 2,273  (139)    & 2,273 (88)    & 2,885 (21,950)  & 0 (0)         & 5,751 (5,751)   & 5,751 (5,751)   & 1 (0)         & 4,436 (4,436)   & 4,436 (4,436)  \\
         {(3)}                                      & 1,511 (56)      & 1,232 (355)        & 1,232  (139)    & 1,232 (56)    & 1,511 (355)    & 0 (0)         & 5,751 (5,751)   & 5,751 (5,751)   & 1 (0)         & 4,436 (4,436)   & 4,436 (4,436)  \\
         {(4)}                          & 3               & 2                  & 3              & 2            & 3             & 0             & 499         & 730          & 0             & 507         & 756    \\
         {(5)}                                 & 6,128           & 2,903              & 6,128           & 2,903         & 6,128          & 6,128          & 6,128        & 6,128         & 2,903          & 2,903        & 2,903   \\
         {(6)}                              & 2               & 1                  & 2              & 1            & 2             & 0             & 54          & 243          & 0             & 16          & 108    \\
         {(7)}                                       & 2               & 1                  & 1              & 1            & 2             & 0             & 7           & 243          & 0             & 11          & 108    \\
         {(8)}                        & 2               & 1                  & 1              & 1            & 2             & 0             & 6           & 243          & 0             & 9           & 108    \\
\end{tabular}
}
\caption{Legitimate calltargets per callsite for each of the eight CFI policies for NodeJS after each CFI defense was individually applied. The values not contained in round brackets are obtained for only virtual callsites and all targets (\textit{i.e.,} virtual and non-virtual), while
the values in round brackets are obtained for all indirect callsites (\textit{i.e.,} virtual and function pointer based calls) and all targets.
For the \textit{Bin types}, \textit{Safe src types}, and \textit{Src types} policies depicted  above the targets can be virtual or non-virtual, for the remaining policies the 
targets inherently can only be virtual functions.
Targets median: (minimum and maximum) number of legal function targets per callsite. Target distribution: minimum/90th percentile/maximum number of targets per callsite. This 90p is determined by sorting the values in ascending order, and picking the value at 90\%. This means that 90\% of the sorted values have a lower or equal value to 90p. P: Policy (Static target constraints), (1) Bin types~\cite{veen:typearmor}, (2) Safe src types~\cite{vtv:tice}, (3) Src types~\cite{mcfi:niu}, (4) Strict src types~\cite{zhang:vtrust}, (5) All virtual tables~\cite{vtint:zhang}, (6) virtual Table hierarchy~\cite{marx:tool}, (7) Sub-hierarchy~\cite{ivt}, 
and (8) Strict sub-hierarchy~\cite{shrinkwrap}.
}
\label{Static target constraints.}
\end{table*}
\begin{table*}[ht]
\centering
\resizebox{2.05\columnwidth}{!}{
  \begin{tabular}{ r|r|r|r|r|r|r|r|r|r|r|r|r } \hline
                                                                 &&    \multicolumn{1}{c|}{\hspace{0.1cm}\textbf{Callsites}} &\multicolumn{2}{c|}{\hspace{0.2cm}\textbf{Targets Baseline}}    & \multicolumn{8}{c}{\textbf{Virtual Function Targets}} \\
         \textbf{\textit{P}}                               &\textit{Value}       &\textit{Write cons.}  &Base. all func.     &Base. vFunc.     & \colorbox{light-gray}{\textit{(1)}}    & \colorbox{light-gray}{\textit{(2)}} & \colorbox{light-gray}{\textit{(3)}} & \colorbox{light-gray}{\textit{(4)}} & \colorbox{light-gray}{\textit{(5)}} & \colorbox{light-gray}{\textit{(6)}}        & \colorbox{light-gray}{\textit{(7)}}  & \colorbox{light-gray}{\textit{(8)}}          \\
\hline

	        &Min	&	&	&	&12,545 (1,956)	&0 (0)	        &0 (0)	        &0 (0)	        &6,128	&0	&0	&0	\\
	        &90p	&	&	&	&30,179 (4,078)	&5,751 (810)	&5,751 (810)	&499 (10)	&6,128	&54	&7	&6	\\
NJS	        &Max	&none	&32,478	&6,300	&32,478 (4,455)	&5,751 (810)	&5,751 (810)	&730 (243)	&6,128	&243	&243	&243	\\
	        &Median	&	&	&	&21,950 (3,106)	&2,885 (426)	&1,511 (121)	&3 (3)	        &6,128	&2	&2	&2	\\
	        &Avg	&	&	&	&19,395 (2,793)	&2,406 (414)	&2,113 (354)	&86 (12)	&6,128	&14	&8	&8	\\

\hline

	        &Min	&	&	&	&2,608 (232)	 &1 (0)	        &1 (0)	        &0 (0)	        &788	&0	&0	&0	\\
	        &90p	&	&	&	&4,085 (546)	 &1,315 (97)	&1,315 (97)	&17 (13)	&788	&34	&7	&7	\\
TS	&Max	&none	&6,201	&796    &6,201 (710)	 &1,315 (159)	&1,315 (159)	&18 (16)        &788	&42	&18	&18	\\
	        &Median	&	&	&	&2,608 (232)	 &1,315 (97)	&1,315 (97)	&17 (13)	&788	&7	&1	&1	\\
	        &Avg	&	&	&	&3,122 (321)     &928 (76)	&923 (74)	&11 (9)	        &788	&10	&3	&3	\\   
	
\hline

	        &Min	&	&	&	&97,041 (37,873)	&0 (0)	             &0 (0)	         &0 (0)	        &68,560	&0	&0	&0	\\
	        &90p	&	&	&	&201,477 (63,816)	&64,315 (24,661)       &64,315 (24,661)	 &48 (30)	&68,560	&192	&25	&15	\\
C        &Max	&none	&232,593	&78,992	&232,593 (71,000)	&64,315 (24,661)	     &64,315 (24,661)	 &3,029 (509)	&68,560  &4,486	&4,486	&4,486	\\
	        &Median	&	&	&	&97,041 (37,873)	&8,672 (4,593)         &7,633 (4,593)	 &3 (2)	        &68,560	&6	&2	&2	\\
	        &Avg	&	&	&	&128,452 (45,731)	&29,315 (11,119)       &29,127 (11,013)      &57 (19)	&68,560	&78	&37	&32	\\

\end{tabular}
}
\caption{Legitmate calltargets per callsite for only virtual callsites and for only the C++ programs after each CFI defense was individually applied.
\textit{Baseline all func.} represents the total number of functions,
while \textit{Baseline virtual func.} represents the number of virtual functions.
The first four policies, from left to right in italic font (\textit{Bin types}, \textit{Safe src types}, \textit{Src types}, and \textit{Strict src types})
allow virtual or non-virtual targets, while the remaining four policies 
inherently allow only virtual targets. This is not a limitation of \sysname but rather how these were intended, designed and used in the original tools from where these are stemming.
The values in round brackets show the theoretical results after adapting the 
first four policies to only allow virtual targets.
Each table entry contains five aggregate values: minimal, 90p: minimum/90th percentile/maximum,
maximal, median and average (Avg) number of targets per callsite.
P: program, NJS: NodeJS, TS: Traffic Server, C: Chrome. (1) Bin types, (2) Safe src types, (3) Src types, (4) Strict src types, (5) All virtual tables, (6) virtual Table hierarchy, (7) Sub-hierarchy, 
and (8) Strict sub-hierarchy.
}
\label{Overall results vcals}
\end{table*}

\begin{table*}[ht!]
\centering
\resizebox{2.05\columnwidth}{!}{%
 \begin{tabular}{ r | r | r | r | r | r | r | r | r | r | r | r | r | r | r | r | r | r | r | r | r | r | r | r | r | r} \hline
         \multicolumn{1}{c|}{\textbf{P}}                                        & \textbf{B}& \multicolumn{3}{c|}{\colorbox{light-gray}{{Bin types}}}     & \multicolumn{3}{c|}{\colorbox{light-gray}{{Safe src types}}}   & \multicolumn{3}{c|}{\colorbox{light-gray}{{Src types}}}       & \multicolumn{3}{c|}{\colorbox{light-gray}{{Strict src types}}}  & \multicolumn{3}{c|}{\colorbox{light-gray}{{All vTables}}}     & \multicolumn{3}{c|}{\colorbox{light-gray}{{vTable hierarchy}}}  & \multicolumn{3}{c|}{\colorbox{light-gray}{{Sub-hierarchy}}}   & \multicolumn{3}{c}{\colorbox{light-gray}{{Strict sub-hierarchy}}}  \\
         \textbf{}                                             &   & \textit{Avg}& \textit{SD}   & \textit{90p}    & \textit{Avg}& \textit{SD} & \textit{90p}      & \textit{Avg}     & \textit{SD} & \textit{90p}& \textit{Avg}     & \textit{SD} & \textit{90p}   & \textit{Avg}     & \textit{SD} & \textit{90p} & \textit{Avg}      & \textit{SD} & \textit{90p}& \textit{Avg} & \textit{SD} & \textit{90p}    & \textit{Avg} & \textit{SD} & \textit{90p}  \\ \hline
NJS		&6,300	&59.72	&21.0	&92.92	&7.41	&6.32	&17.71	&6.51	&6.44	&17.71	&0.26	&0.54	&1.54	&97.27	&0.0	&97.27	&0.23	&0.63	&0.86	&0.13	&0.46	&0.11	&0.13	&0.46	&0.1	\\
TS	&796	&50.35	&15.79	&65.88	&14.97	&8.89	&21.21	&14.89	&9.01	&21.21	&0.18	&0.12	&0.27	&98.99	&0.0	&98.99	&1.26	&1.27	&4.27	&0.34	&0.51	&0.88	&0.34	&0.51	&0.88	\\
C	&78,992	&55.23	&19.08	&86.62	&12.6	&12.16	&27.65	&12.52	&12.22	&27.65	&0.02	&0.11	&0.02	&86.79	&0.0	&86.79	&0.1	&0.43	&0.24	&0.05	&0.41	&0.03	&0.04	&0.41	&0.02	\\\hline
\textit{Avg.}&28,696  &\colorbox{light-gray}{55.1}   &18.62  &81.8   &\colorbox{light-gray}{11.66}  &9.12   &22.19  &\colorbox{light-gray}{11.3}   &9.22   &22.19  &\colorbox{light-gray}{0.15}   &0.25   &0.61   &\colorbox{light-gray}{94.35}  &0.0    &94.35  &\colorbox{light-gray}{0.53}   &0.77   &1.79   &\colorbox{light-gray}{0.17}   &0.46   &0.34   &\colorbox{light-gray}{0.17}   &0.46   &0.33   \\
\end{tabular}
}

\caption{Normalized results with the \textbf{Baseline} using only virtual callsites. Note that virtual callsites all eight policies can be used as these were designed in the original papers to be used for these types of callsites as well.
Baseline: Total number of possible virtual targets.
Each entry contains three aggregate values: average-, standard deviation (SD)
 and 90p-number of targets per callsite. The lower the \textit{Average} value is the better the CFI defense is. B: baseline, P: program, NJS: NodeJS, TS: Traffic Server, C: Chrome}
\label{Standard dev. for virtual tagerts}
\end{table*}

In this Section, we show \sysname usefulness by addressing the following research questions (RQs):
\textbf{RQ1:} What is the residual attack surface of NodeJS (we performed an use case) after eight state-of-the-art CFI policies are independently applied (\cref{rq1})? 
\textbf{RQ2:} What score would each of the analyzed CFI defenses get  (\cref{rq9})?
\textbf{RQ3:} How can \sysname be used to rank CFI policies based on the offered protection level (\cref{rq3})? 
\textbf{RQ4:} What is the residual attack surface for several real-world analyzed programs (\cref{rq2})?
\textbf{RQ5:} How can \sysname be used to construct code reuse attacks? (\cref{rq7})?

\textbf{Test Programs.} 
In our evaluation, we used the following real-world programs:
Nginx \cite{nginx} (Web server, usable also as: reverse proxy, load balancer, mail proxy and HTTP cache, v.1.13.7, \verb!C! code), 
NodeJS \cite{nodejs} (cross-platform JavaScript run-time environment, v.8.9.1, \verb!C/C++! code),
Lighttpd \cite{lighthttpd} (Web server optimized for speed-critical environments, v.1.4.48, \verb!C! code), 
Httpd \cite{httpd} (cross-platform Web server, v.2.4.29, \verb!C! code), 
Redis \cite{redis} (in-memory database with in-memory key-value store, v.4.0.2, \verb!C! code), 
Memcached \cite{memcached} (general-purpose distributed memory caching system, v.1.5.3, \verb!C/C++! code), 
Apache Traffic Server \cite{trafficserver} (modular, high-performance reverse proxy and forward proxy server, v.2.4.29, \verb!C/C++! code), and 
Chrome \cite{chrome} (Google's Web browser, v.33.01750.112, \verb!C/C++! code).

\textbf{Experimental Setup.} The experiments were performed on an Intel i5-3470 CPU with 8GB of RAM running on the Linux Mint 18.3 OS. All experiments were
performed ten times to provide reliable values. If not otherwise stated, we modeled each of the eight CFI defenses inside \sysname according to the policy descriptions provided in Section \ref{Mapping Defenses Into CFI-Assessor}.

\textbf{Reckon's Metric.}
Let $ics_{i}$ be a particular indirect callsite in a program $P$,
$ctr_{i}$ is the total number of legitimate calltargets for an $ics_{i}$ 
after hardening a program with a certain CFI policy.
Then the \textit{CTR} metric is: $CTR = \sum_{i=1}^{n} ctr_{i}$.
Note that the lower the value of $CTR$ is for a given program, the more precise the CFI policy is.  We used this metric to compute the residual attack surface after a CFI defense was applied.

\subsection{Detailed Analysis of NodeJS}
\label{rq1}
In this Section, we analyze the residual attack surface after each of the eight CFI policies 
was applied individually to NodeJS. Note that three out of the eight assessed CFI policies used in the following Tables are the same 
as used by Veen \textit{et al.} \cite{newton} (we share the same names). For the other five CFI policies we use names which reflect their particularities.
We selected NodeJS as this is a very popular real-world application and it contains both \verb!C! and \verb!C++! code. As such, \sysname
can collect results for the \verb!C! and \verb!C++! related CFI polices.

\autoref{Static target constraints.} depicts the static target constraints for the NodeJS 
program under different static CFI calltarget constraining policies. \autoref{Static target constraints.} provides the minimal and
maximum values of virtual calltargets which are available for a virtual callsite after one of the eight 
CFI policies is applied. MKSnapShot contains the Chrome V8 engine and is
used as a shared library by NodeJS after compilation. We decided to add MKSnapshot in \autoref{Static target constraints.} as this component 
is strongly used by NodeJS and represents a source of potential calltargets. The NodeJS results were obtained after static linking of MKSnaphot.

The targets median entries in \autoref{Static target constraints.} (left hand side) indicate the median values obtained for independently
applying one of the eight CFI policies to NodeJS. For both NodeJS and MKSnaphot, the best median number 
of residual targets is obtained using the following policies:
(1) \textit{vTable hierarchy}, 
(2) \textit{Sub-hierarchy}, and
(3) \textit{Strict sub-hierarchy}. These results indicate that these three CFI policies provide the lowest attack surface
while the highest attack surface is obtained for the \textit{Bin types} policy, which allows the highest number of virtual and non-virtual targets. 

The targets distribution in \autoref{Static target constraints.} (right hand side) shows the minimum, maximum and 90 percentile results
for the same eight policies as before. While the minimum value is
0 the highest values for both NodeJS and MKSnapshot are obtained for the \textit{Bin types} policy, while the lowest values
are obtained for the following policies:
(1) \textit{vTable hierarchy}, 
(2) \textit{Sub-hierarchy}, and
(3) \textit{Strict sub-hierarchy}.
Further, the 90p results show that on the tail end of the distribution, a noticeable difference between the three previously mentioned 
policies exists. We can observe that for these critical callsites the \textit{Strict sub-hierarchy} policy provide the least amount of residual 
targets and therefore the best protection against CRAs. Meanwhile, the 90p results for the \textit{Strict src type} 
and \textit{vTable hiearchy} policies 
indicate that the residual attack surface might still be sufficiently large for the attacker. 

\subsection{CFI Defenses Scores}
\label{rq9}

\begin{figure}[ht]
\centering
  \includegraphics[width=0.92\columnwidth]{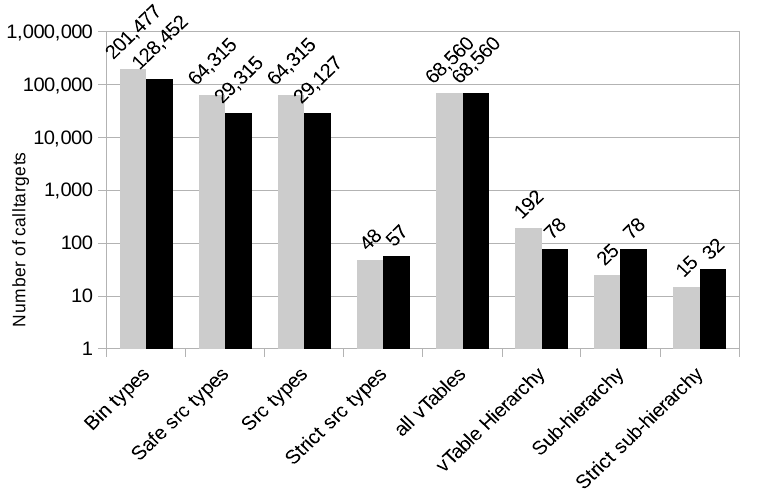}
	\caption{Scores obtained by each analyzed CFI defense. The scores are depicted in logarithmic scale in order to better compare the values against each other. The optimal score has the value of one depicted on the left had side $Y$ axis.}
	\label{scores}
\end{figure}

\autoref{scores} depicts the scores obtained by each of the eight policies which were analyzed for the Chrome Web browser. We opted to depict the values for only the Chrome browser since this represents the largest (approx. 10 Mil. LOC) analyzed program. The numbers on the gray shaded bars represent the 90p values while the values on the black shaded bars represent the average values for the Chrome Web browser. These values can be observed in \autoref{Overall results vcals} on the last row from top to bottom for the Chrome browser as well. The optimal score is one and means that each callsite is allowed to target a single calltarget. This is the case only during runtime. The lower the bar is or the closer the value is to one the better the score is.  \looseness=-1 

The best score w.r.t. the 90p and average values is obtained for the \textit{Strict sub-hiearchy} which is the best CFI defense from the eight analyzed policies. Interestingly to note that the best function signature based policy \textit{Stric src types} has a slightly worse score than the second based class based CFI defense (\textit{Sub-hierarchy}). Finally, we note that these CFI-based forward edge policies are not optimal (provide values larger than one)
and the desired goal is to have such policies which provide a one-to-one mapping as shadow stack based techniques.

\subsection{Ranking of CFI Policies}
\label{rq3}
In this Section, we normalize the results presented in RQ2 using the \textit{Baseline} values, (\textit{i.e.,} the number of possible target functions), in order to be able to compare the assessed CFI policies against each other w.r.t. calltarget reduction.
This allows to compare the analyzed CFI defenses on programs with different sizes and complexities which would be not possible otherwise.

\autoref{Standard dev. for virtual tagerts} depicts the average, standard deviation and 90th percentile results 
obtained after analyzing only virtual callsites. Unless stated otherwise, we use the $CTR$ metric introduced at the beginning of this section.
For these callsites, all eight CFI policies can be assessed.

We calculated the average over the three \verb!C++! programs after normalization.  
By considering these aggregate average values, the eight policies 
can be ranked (from best (smallest aggregate average) to worst (highest aggregate average)) as follows:
(1) \textit{Strict src types} (0.15),
(2) \textit{Strict sub-hierarchy} (0.17),
(3) \textit{Sub-hierarchy} (0.17),
(4) \textit{vTable hierarchy} (0.53),
(5) \textit{Src types} (11.3),
(6) \textit{Safe src types} (11.66),
(7) \textit{Bin types} (55.1), and 
(8) \textit{All vTables} (94.35). 

From the class hierarchy-based policies \textit{Strict sub-hierarchy} perform best in all three 
aggregate results (Avg, SD and 90p). In comparison, \textit{Strict sub-hierarchy} performs better 
w.r.t. average and standard deviation but worse w.r.t. 90p. 
These results indicate that these two policies are the most restrictive, but
a clear winner in all evaluated criteria cannot be determined.

\begin{table}[ht]
\centering
\resizebox{1.0\columnwidth}{!}{%
 \begin{tabular}{ r | r | r | r | r | r | r | r | r | r | r } \hline
         \multicolumn{1}{c|}{\textbf{}}                                & &\multicolumn{3}{c|}{\hspace{0.0cm}\textbf{Bin types}}           & \multicolumn{3}{c|}{\hspace{0.0cm}\textbf{Safe src types}}     & \multicolumn{3}{c}{\hspace{0.0cm}\textbf{Src types}}    \\
         \textbf{P}                                              &\textit{B}  & \textit{Avg}& \textit{SD}&\textit{90p}     & \textit{Avg}& \textit{SD}&\textit{90p}          &\textit{Avg} & \textit{SD}&\textit{90p}    \\ \hline
a		&32,478	&64.0	&20.43	&92.92	&3.82	&5.83	&17.71	&3.38	&5.64	&17.71	\\
b	&6,201	&54.03	&18.76	&87.89	&13.54	&9.27	&21.21	&13.46	&9.36	&21.21	\\
c	&232,593	&56.83	&19.84	&86.62	&11.71	&12.11	&27.65	&11.64	&12.16	&27.65	\\
d		&1,949	&52.18	&26.5	&92.0	&2.7	&3.01	&8.21	&2.46	&3.01	&8.21	\\
e	&594	&65.25	&27.81	&97.98	&2.94	&3.18	&7.41	&2.93	&3.19	&7.41	\\
f	&225	&69.75	&7.11	&68.89	&1.0	&0.97	&0.89	&1.0	&0.97	&0.89	\\
g		&1,270	&54.91	&24.85	&92.28	&6.38	&4.56	&11.73	&6.36	&4.57	&11.73	\\
h		&2,880	&65.19	&16.51	&84.62	&1.25	&2.52	&1.88	&1.2	&2.52	&1.88	\\\hline
\textit{Avg}&34,773	&\colorbox{light-gray}{60.3}	&34.39	&87.9	&\colorbox{light-gray}{5.4}	&5.18	&12.09	&\colorbox{light-gray}{5.3}	&5.17	&12.08	\\
\end{tabular}
}
\caption{Normalized results using all indirect callsites. B:baseline, a:NodeJS, b:Apache Traffic Server,
c:Google's Chrome, d:Httpd, e:LightHttpd, f:Memcached, g:Nginx, h:Redis.}
\label{Average and standard deviation for not virtual targets}
\end{table}
\autoref{Average and standard deviation for not virtual targets} depicts (similarly to \autoref{Standard dev. for virtual tagerts})
normalized results with the difference that all indirect callsites (both virtual and pointer based) are analyzed.
Thus, the \textit{Baseline} values used for normalization include virtual and non-virtual targets. 
By taking into account the aggregate averages and the standard deviation of the three policies
in \autoref{Average and standard deviation for not virtual targets},
we can rank the policies as follows (from best to worse):
(1) \textit{Src types} (Avg 5.3 and SD 5.17),
(2) \textit{Safe src types} (Avg 5.4 and SD 5.18), and 
(3) \textit{Bin types} (Avg 60.3 and SD 34.39).

By considering the 90p values we conclude that 
for the most vulnerable 10\% of callsites, \textit{Bin types} only restricts the target set to 
87.9\% of the unprotected target set. These callsites essentially remain unprotected.
Meanwhile, the \textit{Safe src type} and \textit{Src type} policies restrict
to only around 12\% of the unprotected target set.

\subsection{Generalized Results}
\label{rq2}
\begin{table}[ht]
\centering
\resizebox{.97\columnwidth}{!}{%
 \begin{tabular}{ p{.1cm} | r | p{.7cm} | p{.99cm} | p{1.4cm} |  p{1.2cm} | p{.9cm}} \hline
                                                                 &&\multicolumn{1}{c|}{\hspace{0.1cm}\textbf{}} &    & \multicolumn{3}{c}{\textbf{Targets (Non-)vFunctions}}          \\
         \textbf{\textit{P}}                               &\textit{Value}       &\textit{Callsite write cons.}  &\textit{Baseline all func.}     & \textit{(1)}    & \textit{(2)} & \textit{(3)}     \\\hline

\hline

	&Min	&	&	&885	        &0	 &0	\\
	&90p	&	&	&30,179	        &5,751	 &5,751	\\
a	&Max	&none	&32,478	&32,478	        &5,751	 &5,751	\\
	&Med	&	&	&21,950	        &88	 &56	\\
	&Avg	&	&	&20,787   	&1,242   &1,099	\\

\hline

	   &Min	        &	&	&357	&0	&0	\\
	   &90p	        &	&	&5,450	&1,315	&1,315	\\
b    &Max	        &none	&6,201	&6,201	&1,315	&1,315	\\
	   &Med	&	&	&2,608	&1,315	&1,315	\\
	   &Avg	&	&	&3,350	&840	&835	\\

\hline

	 &Min	&	&	&3,612	        &0	  &0	\\
	 &90p	&	&	&201,477	&64,315	  &64,315	\\
c    &Max	&none	&232,593&232,593	&64,315	  &64,315	\\
	 &Med&	&	&97,041	        &8,672	  &7,394	\\
	 &Avg	&	&	&132,182  	&27,238   &27,074	\\

\hline

	&Min	&	&	&99	        &0	&0	\\
	&90p	&	&	&1,793	        &160	&160	\\
d	&Max	&none	&1,949	&1,915	        &160	&160	\\
	&Med	&	&	&1,070	        &18	&16	\\
	&Avg	&	&	&1,017    	&53	&48	\\

\hline

	&Min	&	&	&37	&0	&0	\\
	&90p	&	&	&582	&44	&44	\\
e   &Max	&none	&594	&582	&44	&44	\\
	&Med	&	&	&395	&6	&6	\\
	&Avg	&	&	&388	&17	&17	\\

\hline

	  &Min	  &	&	&92	&0	&0	\\
	  &90p	  &	&	&155	&2	&2	\\
f     &Max	  &none	&225	&221	&17	&17	\\
	  &Med &	&	&155	&2	&2	\\
	  &Avg	  &	&	&157	&2	&2	\\

\hline

	&Min	&	&	&422	&1	&1	\\
	&90p	&	&	&1,172	&149	&149	\\
g	&Max	&none	&1,270	&1,259	&149	&149	\\
	&Med	&	&	&719	&75	&75	\\
	&Avg	&	&	&697	&81	&81	\\

\hline

	&Min	&	&	&1,266	        &1	&1	\\
	&90p	&	&	&2,437	        &54	&54	\\
h	&Max	&none	&2,880	&2,635	        &391	&391	\\
	&Med	&	&	&1,994	        &16	&14	\\
	&Avg	&	&	&1,877   	&36	&35	\\

\hline                                               
\end{tabular}
}
\caption{Results for virtual and pointer based callsites. P: program, Memchd: Memcached, (1) Bin types, (2) Safe src types, (3) Src types. a:NodeJS, b:Apache Traffic Server,
c:Google's Chrome, d:Httpd, e:LightHttpd, f:Memcached, g:Nginx, h:Redis.}
\label{Overall results function pointer}
\end{table}

\autoref{Overall results function pointer} provides
results for the three policies which can provide protection for both \verb!C! and \verb!C++! programs.
In contrast to \autoref{Overall results vcals}, 
all indirect calls are taken into account (including virtual calls).
Therefore, the targets can be virtual or non-virtual.
Intuitively, the residual attack surface grows with the size of the program. This can be observed 
by comparing the results for large (\textit{e.g.,} Chrome) with smaller (\textit{e.g.,} Memcached) programs.

\autoref{Overall results vcals} depicts the overall results obtained after applying the eight assessed CFI policies to virtual callsites only.
The first four policies (italic font)
cannot differentiate between virtual and non-virtual calltargets. 
Therefore, for these policies 
the baseline of possible calltargets includes all functions 
(both virtual and non-virtual). This is denoted with Baseline all func.
Since the remaining four policies can only be applied to virtual callsites, they restrict the possible calltargets to only virtual functions.
Thus, the baseline for these policies includes only virtual functions (Baseline virtual function). 
For a better comparison between the first and second categories of policies,
we also calculated the target set when restricting the first four policies
to only allow virtual callsites.
For \textit{Bin types}, \textit{Safe src types}, \textit{Src types}, and \textit{All vTables} the results indicate that there is \textit{no} protection offered.
The three-class hierarchy-based policies perform best when considering the median and average results. In addition, the 
\textit{Strict src type} policy performs surprisingly well, especially after restricting the target set to only virtual functions. Med: median.

\subsection{Towards Automated CRAs Construction}
\label{rq7}
In this Section, we show how \sysname is used to build an attack which bypasses a state-of-the-art CFI policy-based defense, namely VTV's \textit{Sub-hierarchy Policy}.
This case study is architecture independent, since \sysname's analysis is performed at the IR level during LTO time in LLVM.
Note that LLVM IR code represents a higher level representation of machine code (metadata), thus
our results can be applied to other architectures (\textit{e.g.,} ARM) as well.
Our case study assumes an ideal implementation of VTV/IFCC. Breaking
the ideal instrumentation shows that the defense can be bypassed in any
implementation.

In this case study, we 
present the required
components for a COOP attack by studying the original COOP attack against Mozilla's Firefox web browser and 
demonstrate
that such an attack is easier to perform when using \sysname. Thus, we discuss the importance of the CRA construction with \sysname at hand.

The original COOP attack presented by Schuster \textit{et al.} \cite{coop} used:
(1) a buffer overflow filled with six fake counterfeit objects by the attacker,
(2) precise knowledge of the Firefox \texttt{libxul.so} layout,
(3) where an COOP dispatcher gadget (\texttt{ML-G}) resides, and
(4) several other useful gadgets in order to open an Unix shell.

In general terms, to pursue a CRA the attacker needs: 
(1) an exploitable memory corruption,
(2) attack starting point (\textit{i.e.,} callsite) becomes corruptible due to (1),
(3) program binary memory layout leak,
(4) appropriate (usable and available) gadgets,
(5) the possibility to read and write into memory,
(6) the possibility to link gadgets and pass information from one to each other, and 
(7) the possibility to perform calls into the system \texttt{stdlib} or any other reach functionality library.

As demonstrated, the attacker first has to find an exploitable memory corruption (\textit{e.g.,} buffer overflow, etc.) and fill it with fake objects.
Next, the attacker calls different gadgets (virtual functions) located in the program binary. 
As such, we assume that NodeJS contains an exploitable memory vulnerability (\textit{i.e.,} buffer overflow), and that the attacker 
is aware of the layout of the program binary. The attacker would then want to bend the control flow to only
per callsite legitimate 
calltargets since he does not know if other defenses are in place. He would also want to avoid calling into other program
class hierarchies. Therefore, he needs to know 
which calltargets are legitimate
for all callsites in the main NodeJS binary.

\begin{table}[ht]
\centering
\resizebox{\columnwidth}{!}{%
 \begin{tabular}{ p{.1cm} | p{.1cm} | p{.8cm} |  p{.9cm} | p{.6cm} | p{.4cm} | p{.4cm} | p{.2cm} | p{.5cm} | p{.2cm} | p{.2cm} | p{.2cm}} \hline
         \multicolumn{1}{c|}{\hspace{.01cm}\textbf{\textbf{CS}}} &\multicolumn{1}{c|}{}  &\multicolumn{1}{c|}{} &\multicolumn{1}{c|}{}                         & \multicolumn{8}{c}{\hspace{0cm}\textbf{Target Policies}} \\
         \textbf{\hspace{1cm}\textbf{}}  & \textit{\#}         &\textbf{Baseline only vFunc.}    &\textbf{Baseline all Func.}          & \textbf{(1)}  & \textbf{(2)} & \textbf{(3)} & \textbf{(4)} & \textbf{(5)}  & \textbf{(6)}        & \textbf{(7)}  &\textbf{(8)}    \\ \hline
         a                     & 5                   & 6,300                           & 32,478                              & 31,305              &4                        &4                   &1                           &6,128                  &1                                 &1                        &1                                 \\
         b                      & 2                   & 6,300                           & 32,478                              & 21,950              &719                      &719                 &49                          &6,128                  &57                                &53                       &49                                \\
         c                        & 3                   & 6,300                           & 32,478                              & 27,823              &136                      &136                 &1                           &6,128                  &1                                 &1                        &1                                 \\
         d                    & 1                   & 6,300                           & 32,478                              & 12,545              &810                      &810                 &1                           &6,128                  &72                                &12                       &12                                \\
         e                     & 1                   & 6,300                           & 32,478                              & 1,956               &810                      &810                 &1                           &6,128                  &72                                &13                       &13                                \\
         f                   & 1                   & 6,300                           & 32,478                              & 1,956               &810                      &810                 &6                           &6,128                  &20                                &19                       &19                                \\
         g                & 3                   & 6,300                           & 32,478                              & 1,956               &810                      &810                 &6                           &6,128                  &20                                &19                       &19                                \\
         h                      & 2                   & 6,300                           & 32,478                              & 3,106               &35                       &35                  &8                           &6,128                  &48                                &13                       &5                                 \\
         i                      & 2                   & 6,300                           & 32,478                              & 3,106               &2,984                    &2,984               &49                          &6,128                  &53                                &53                       &49                                \\
         j                      & 2                   & 6,300                           & 32,478                              & 3,106               &719                      &719                 &49                          &6,128                  &53                                &53                       &19                                \\
\end{tabular}
}
\caption{Ten controllable callsites and their legitimate targets under the \textit{Sub-hierarchy} CFI defense. \#: passed parameters. CS: Ten controllable callsites, (1) Bin types, (2) Safe src types, (3) Src types, (4) Strict src types, (5) All virtual tables, (6) virtual Table hierarchy, (7) Sub-hierarchy, 
and (8) Strict sub-hierarchy.}
\label{Controllable callsites.}
\end{table}

\autoref{Controllable callsites.} depicts ten controllable callsites (in total \sysname found thousands of controllable callsies) for which 
the legitimate target set, depending on the used CFI policy, ranges from one to \numprint{31,305} calltargets. a:debugger.cpp:13 29:33, b:protocol.cpp:839:60, c:schema.cpp:133:33, d:handle\_wrap.cc:127:3, e:cares\_wrap.cc:642:5, f:node\_platform.cc:25:5, g:node\_http2 \_core.h:417:5, h:tls\_wrap.cc:771:10, i:protocol.cpp:839:60, and j:protocol.cpp:836:60.

For each calltarget, \sysname provides: file name, function name, start address and source code line number such that it can be easily 
traced back in the source code file.
The calltargets (right hand side in \autoref{Controllable callsites.} in
italic font) represent available calltargets for each of the eight assessed
policies.

For our case study, we decided not to use the most restrictive CFI policy (\textit{i.e.,} \textit{Strict sub-hierarchy})
as it enforces the same function in the virtual table sub-hierarchy. As such, only through inheritance, chances are 
that the implementation of the function may vary from the initial implementation. Thus, the number of useful gadgets would be low.

Consequently, we assume that the attacker knows that the NodeJS binary is protected with the 
\textit{Sub-hierarchy} policy. The attacker ranks all callsites depending on the number of usable calltargets. 
By ranking the callsites w.r.t. their residual target set, the attacker wants to know the 
precise functions (calltargets) which are allowed for each callsite. Further, in order to perform the attack, 
he has to access the source code of the application. 
After searching through the source code, he finds several callsites where vector-based object dispatches are performed.
The attacker next finds out that 
the detected calltarget set contains several usable gadgets. 
Note that for the COOP attack to be performable, the only hard constraint is that at least one usable 
\texttt{ML-G} gadget must exist in the program. This constraint was addressed by the attacker at the beginning of their source code search since they decided 
to only look for callsites which are part of an \texttt{ML-G} COOP gadget.
As such, the attacker knows exactly which gadgets are available for each controllable callsite. 
As none of the static CFI policies impose write constraints by default (\textit{i.e.,} none) the attacker can overflow a buffer at a certain callsite with fake 
objects. Further, they start to call their gadgets one by one and passing information over the stack through scratch registers from one gadget to the next one.
Note that their attack does not violate the CFI check in place at the selected object dispatch location since this is contained in a \texttt{ML-G} gadget.
Further, he can deliberately avoid violating the original program control flow by calling into other program class hierarchies or other illegitimate 
calltargets as the original COOP attack does. Eventually, the attacker succeeds in opening an Unix shell.

Finally, this shows the usefulness of \sysname for an analyst, when searching for legitimate gadgets that available for each callsite after a certain CFI policy was applied. 
Thus, this helps to better tailor his attack w.r.t. the deployed CFI-based defense. 

\section{Discussion}
\label{Discussion}
\textit{\sysname's Metrics vs. other CFI Metrics.}
Existing CFI-defense assessing metrics (\textit{e.g.,} AIR, fAIR and AIA) are designed to reason about CFI based defenses by providing
average values obtained mostly by dividing the total number of control flow transfers to the total number of callistes for example. 
Also, when computing 
these values the ground truth numbers w.r.t. available calltargets, total number of calltargets, total number of unprotected 
calltargets are in most cases not provided. Thus, it is hard for any researcher which looks for reproducibility of these results to compare his 
results with other research results.
Further, these metrics do not reason about other aspects of CFI-based defenses such as the forward check runtime overhead,
return site target reduction, return site target runtime overhead and availability of gadgets at the end of the forward or backward edges.
Thus the only benefit of these metrics are some average numbers which are hard to reproduce because of the above mentioned reasons.
Finally, similar to \sysname these three metrics do not reason about gadget link-ability which would help to shed light in areas of optimal 
CFI-based protection schemes which then could be effectively used to protect applications without the need to enforce a strict CFI-policy on the whole
program binary or libraries. More specifically, think as if the defender could harden a \verb!C++! based application such that all main loop gadgets (and alike) are 
made unusable for any attacker. This would remove a crucial building block for a COOP attack and thus the attacker would not have an important building 
block in his arsenal. Thus, the COOP attack would not be possible.

\sysname's metrics are superior compared to AIR, fAIR and AIA metrics since these allow to quantify static CFI-based defenses w.r.t. to 
more dimensions. Further, \sysname allows to precisely reason about the forward edge based calltarget reduction by allowing
to better compare the obtained results with ground truth numbers, by for example not averaging the results and providing the first framework which allows to comprehensively
reason about a CFI defense. Finally, \sysname can be used to reason quantitatively and consistently about object-oriented programing (OOP) concepts 
which represent the building blocks for many CFI defenses.

\section{Limitations}
\label{Limitations}
{\textit{Evaluation Results.}}
We are aware that our evaluation results reflect the measurements obtained by applying \sysname 
on several real-world open source programs which may not generalizable.
We therefore believe that more programs should be evaluated with the help of \sysname (the first CFI-based 
consistently reasoning framework for static CFI defenses). Thus, we urge other researchers to use \sysname's CFI-based metrics when
assessing their defense and comparing it with other existing tools. Finally, we envisage that \sysname could
become the de-facto-standard benchmarking framework for assessing static CFI-based defenses.



\section{Conclusion}
\label{Conclusions}

We have presented \sysname, a CFI defense assessment and gadget search framework which allows for the first time to thoroughly compare CFI defenses against each other.
We implemented \sysname, on top of the Clang/LLVM compiler framework which offers the possibility to precisely analyze real-world programs.
By using \sysname, an analyst can drastically cut down the time needed to search for gadgets which are compatible with state-of-the-art CFI defenses contained in many real-world programs.
Our experiment results indicate that most of the CFI defenses are too permissive. Further, if an attacker does not
only rely on the program binary when searching for gadgets and has a tool such as \sysname at hand to analyze the 
source code of the vulnerable application, then many CFI defenses can be easily bypassed. Finally, in order to 
support further research we plan to release \sysname's source code as open source upon paper acceptance.  \looseness=-1



\end{document}